\documentclass[prd,twocolumn,showpacs,nofootinbib,amsmath,amssymb,superscriptaddress]{revtex4-1}
\usepackage{graphicx}
\usepackage{dcolumn}
\usepackage{bm}
\usepackage{relsize}
\usepackage{amsmath,amssymb,bbm,mathrsfs,nicefrac}
 \usepackage{array,multirow}
 \usepackage{float}
\usepackage{tikz}
\usepackage{amsmath}
\usepackage{amssymb}
\usepackage{graphicx,textcomp,float,gensymb,wrapfig, enumitem,comment,dsfont,framed,slashed,appendix,wrapfig,xcolor}
\usepackage[utf8]{inputenc}

\begin{document}

%
%
\title{Lattice-QCD-based equations of state at finite temperature and density}
\author{Jamie M. Karthein}
\affiliation{Center for Theoretical Physics, Massachusetts Institute of Technology, Cambridge, MA 02139}

\author{Debora Mroczek}
\affiliation{Illinois Center for Advanced Studies of the Universe, Department of Physics, University of Illinois at Urbana-Champaign, Urbana, IL 61801, USA}

\author{Angel R. Nava Acu\~na}
\affiliation{Physics Department, University of Houston, Houston, TX 77204, USA}

\author{Jacquelyn Noronha-Hostler}
\affiliation{Illinois Center for Advanced Studies of the Universe, Department of Physics, University of Illinois at Urbana-Champaign, Urbana, IL 61801, USA}

\author{Paolo Parotto}
\affiliation{Pennsylvania State University, Department of Physics, University Park, PA 16802, USA}

\author{Damien R. P. Price}
\affiliation{Physics Department, University of Houston, Houston, TX 77204, USA}

\author{Claudia Ratti}
\affiliation{Physics Department, University of Houston, Houston, TX 77204, USA}
%
%

%
%

\date{\today}

\begin{abstract}
\vspace{1em} 
%
%
The equation of state (EoS) of QCD is a crucial input for the modeling of heavy-ion-collision (HIC) and neutron-star-merger systems. 
Calculations of the fundamental theory of QCD, which could yield the true EoS, are hindered by the infamous Fermi sign problem which only allows direct simulations at zero or imaginary baryonic chemical potential. 
As a direct consequence, the current coverage of the QCD phase diagram by lattice simulations is limited. 
In these proceedings, two different equations of state based on first-principle lattice QCD (LQCD) calculations are discussed. 
The first is solely informed by the fundamental theory by utilizing all available diagonal and non-diagonal susceptibilities up to $\mathcal{O}(\mu_B^4)$ in order to reconstruct a full EoS at finite baryon number, electric charge and strangeness chemical potentials.
For the second, we go beyond information from the lattice in order to explore the conjectured phase structure, not yet determined by LQCD methods, to assist the experimental HIC community in their search for the critical point. 
We incorporate critical behavior into this EoS by relying on the principle of universality classes, of which QCD belongs to the 3D Ising Model. 
This allows one to study the effects of a singularity on the thermodynamical quantities that make up the equation of state used for hydrodynamical simulations of HICs. 
Additionally, we ensure that these EoSs are valid for applications to HICs by enforcing conditions of strangeness neutrality and fixed charge-to-baryon-number ratio. 
\vspace{1em}
\end{abstract}


\maketitle

\section{Introduction}

The equation of state (EoS)  is a fundamental thermodynamic relationship between state variables, including temperature, volume, and pressure.
This relationship between the pressure, temperature, and volume is defined via statistical mechanics, where the pressure is given as the logarithm of the partition function.

The full equation of state of QCD is still not known from first-principles lattice simulations due to the limitations of the method arising from the Fermi sign problem.
Currently, the EoS is available for vanishing and small values of the baryon chemical potential \cite{Borsanyi:2010cj,Borsanyi:2013bia,Bazavov:2014pvz,Guenther:2017hnx,Gunther:2017sxn,Bazavov:2017dus,Borsanyi:2021sxv,Mondal:2021jxk}.
Lattice QCD results in this regime cover the same region of the phase diagram as heavy-ion collisions performed at the Large Hadron Collider (LHC) and the top energies of the Beam Energy Scan (BES) program at the Relativistic Heavy-Ion Collider (RHIC).
However, in order to map out the phase structure of QCD across the phase diagram, it is important to extend the equation of state to chemical potentials relevant for the entire range of BES energies, including the second phase, BES-II.
This experimental program covers a range of center-of-mass collision energies per nucleon pair, $\sqrt{s_{\text{NN}}}$, from 200 GeV down to 7.7 GeV, along with fixed target experiments resulting in energies as low as 3.0 GeV.
This corresponds to a range in baryonic chemical potential of up to $\mu_B \leq 420$ MeV, not including the fixed target experiments.

An important theoretical framework for understanding the properties of the medium produced in HICs is hydrodynamic simulations of strongly-interacting matter, which has found striking agreement with experimental data by describing the system as a fluid \cite{Kolb:2000sd,Huovinen:2001cy,Kolb:2002ve,Kolb:2003dz,Heinz:2013th,An:2021wof}.
These hydrodynamic simulations depend on the QCD equation of state as required input.
Therefore, in order to provide an interpretation of the experimental results across BES energies with hydrodynamic simulations, the full EoS needs to be extended to relatively large values of the chemical potential.
In my dissertation research, I have developed several different equations of state to be used in hydrodynamic simulations of heavy-ion collisions in order to provide insight into the exotic, deconfined matter created in these collisions.
The work described in this chapter is based, in part, on previously published research from Refs. \cite{Noronha-Hostler:2019ayj,Parotto:2018pwx,Karthein:2021nxe}.

\section{BQS EoS}
The phase diagram of QCD that is typically considered in the context of heavy-ion collisions is within the temperature and baryonic chemical potential plane.
However, the true nature of QCD lies within a more complex phase diagram that takes into account the dependence on the chemical potentials of all conserved charges in QCD, namely $\mu_B$, $\mu_S$, and $\mu_Q$.
These are the relevant conserved charge chemical potentials to be considered since we know that the strong nuclear force conserves baryon number, electric charge, and strangeness in all interactions.
This four-dimensional hyperplane describing the phase structure of QCD can be constructed via a Taylor expansion of the QCD pressure in all three chemical potentials.
The necessary Taylor expansion coefficients for this procedure come from first-principles lattice QCD data on the susceptibilities of conserved charges \cite{DElia:2016jqh,Bazavov:2012jq,Borsanyi:2018grb}.
All such susceptibilities were calculated by the Wuppertal-Budapest collaboration up to $\mathcal{O}(\mu_B^4)$ in Ref. \cite{Borsanyi:2018grb}.
Given this, we calculated the pressure as a Taylor series of the three chemical potentials, with coefficients taken from lattice simulations, and constructed the QCD equation of state in a range of temperature
and chemical potential
relevant for the BES-II program.

The Taylor series of the pressure in terms of the three conserved charge chemical potentials is written as:

\begin{equation} \label{eq:BQS_pressTaylor}
\frac{P(T,\mu_B,\mu_Q,\mu_S)}{T^4} =  \sum_{i,j,k}\frac{1}{i!j!k!}\chi_{ijk}^{BQS} 
\left(\frac{\mu_B}{T}\right)^i\left(\frac{\mu_Q}{T}\right)^j\left(\frac{\mu_S}{T}\right)^k. \nonumber
\end{equation}
The first term, $i=j=k=0$, corresponds to the pressure itself as calculated on the lattice, and the subsequent Taylor expansion coefficients are the conserved charge susceptibilities with appropriate factorial coefficients corresponding to the order of the susceptibility.
The susceptibilities are simply the derivatives of the QCD pressure with respect to the various conserved charge chemical potentials:
\begin{eqnarray} \label{eq:BQS_susc}
\chi_{ijk}^{BQS}=\left.\frac{\partial^{i+j+k}(P/T^4)}{\partial(\frac{\mu_B}{T})^i\partial(\frac{\mu_Q}{T})^j\partial(\frac{\mu_S}{T})^k}\right|_{\mu_B,\mu_Q,
\mu_S=0}.
\end{eqnarray}

Since the temperature range of the lattice calculations is not enough to cover the entire hydrodynamical evolution of the system from the creation of the hot, dense quark-gluon plasma (QGP) to hadronization, we continued the results as a function of temperature by smoothly merging each coefficient at low temperature to the Hadron Resonance Gas (HRG) model results.
At high temperatures, we imposed a smooth approach to the Stefan-Boltzmann limit.
We were then able to parametrize these coefficients in order to obtain a smooth description of each one over the entire temperature range from $30 \leq T \leq 800$ MeV. 
The parametrization utilized are shown in Eq. \ref{eq:param_chi2B} for the quantity $\chi_2^B$ and in Eq. \ref{eq:param} for all other susceptibilities.
\begin{equation} \label{eq:param_chi2B}
\chi_2^B(T) = e^{-h_1/x^\prime - h_2/{x^\prime}^2} \cdot f_3 \cdot (1 + \tanh(f_4 x^\prime + f_5)).
\end{equation}
In both equations, $x = T/154 \, \text{MeV}$, $x^\prime = T/200 \, \text{MeV}$.

\begin{widetext}
\begin{equation} \label{eq:param}
\chi_{ijk}^{BQS}(T) = \frac{a^i_0 + a^i_1/x + a^i_2/x^2 + a^i_3/x^3 + a^i_4/x^4 + a^i_5/x^5+ a^i_6/x^6+ a^i_7/x^7+ a^i_8/x^8+ a^i_9/x^9}{b^i_0 + b^i_1/x + b^i_2/x^2 + b^i_3/x^3 + b^i_4/x^4 + b^i_5/x^5+b^i_6/x^6+b^i_7/x^7+b^i_8/x^8+b^i_9/x^9} + c_0.
\end{equation}
\end{widetext}

By utilizing the Taylor expansion shown in Eq. \eqref{eq:BQS_pressTaylor} and our parametrized coefficients, we obtained the pressure and calculated all other EoS quantities from thermodynamic relations.

\begin{figure*}[h!]
\centering
\includegraphics[width=\textwidth,height=7cm]{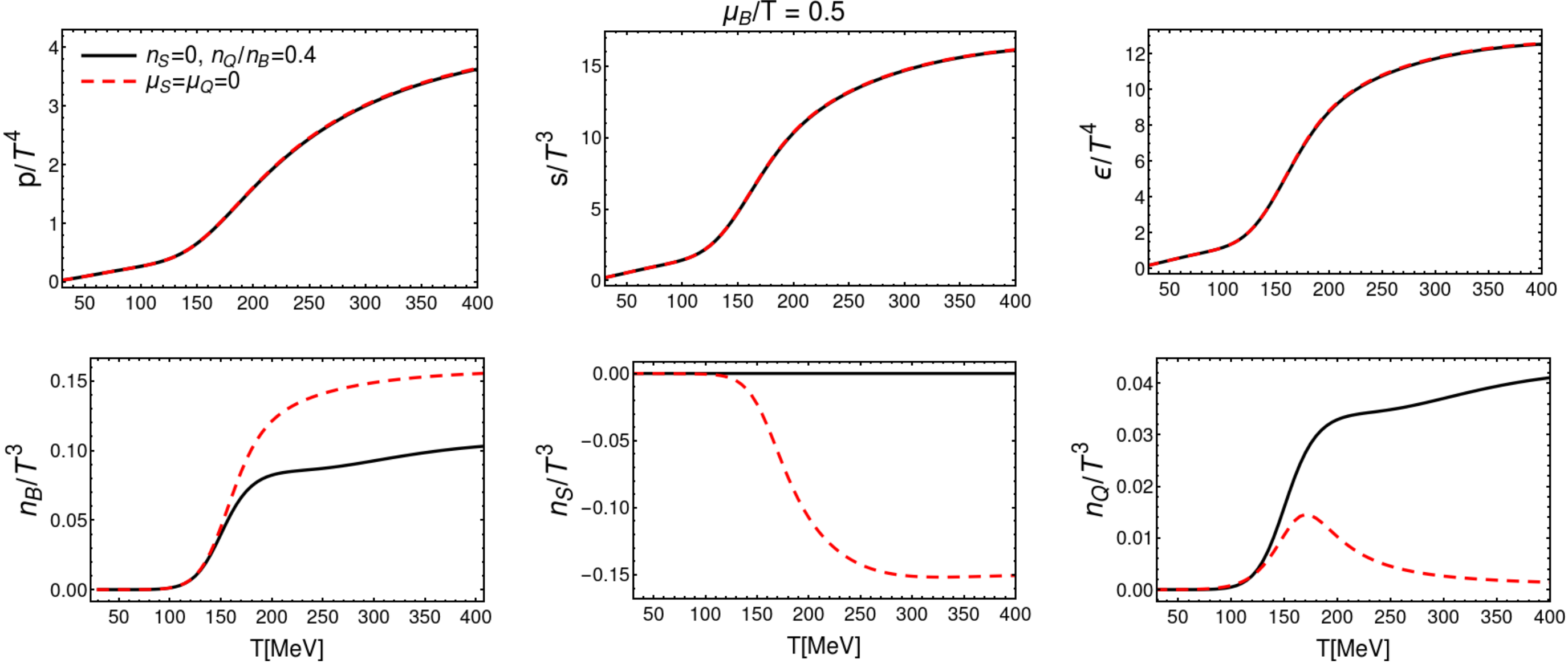} 

\includegraphics[width=\textwidth,height=7cm]{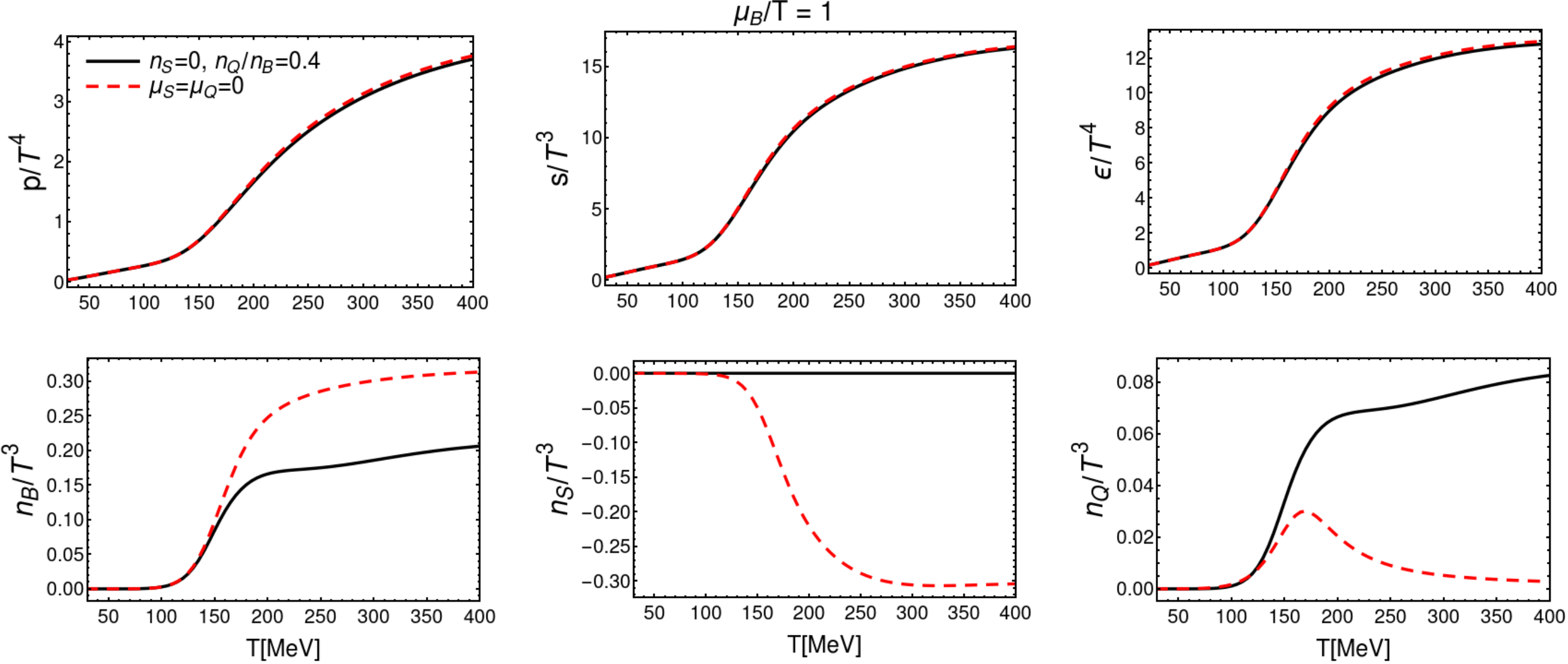} 
\caption{Normalized pressure, entropy density, energy density, baryonic, strangeness and electric charge densities are shown as functions of the temperature along the $\mu_B/T = 0.5$ (top panel), $\mu_B/T = 1.0$ (bottom panel). In all plots, the solid black curves indicate the case $\langle n_S\rangle=0$ and $\langle n_Q\rangle=0.4\langle n_B\rangle$, whereas the dashed red ones indicate the case $\mu_S=\mu_Q=0$ \cite{Noronha-Hostler:2019ayj}.}
\label{fig3}
\end{figure*}

In Fig. \ref{fig3} , the dependence of the normalized pressure, entropy density, energy density, baryonic, strangeness, and electric charge densities on the temperature is shown, along lines of constant $\mu_B/T=0.5,~1$.
These curves were calculated for two choices of the chemical potential constraints.
The first is phenomenologically relevant for HICs, where $\langle n_S\rangle=0$, $\langle n_Q\rangle=0.4\langle n_B\rangle$ (solid black lines). 
These conditions of strangeness neutrality and fixed baryon-to-electric-charge densities are the conditions realized by the experimental situation.
The other choice was given for $\mu_S=\mu_Q=0$ (dashed red lines), which only considers the dependence of the EoS on $T ~ \text{and} ~\mu_B$.

The thermodynamic quantities that are less sensitive to the chemical composition of the system, namely the pressure, entropy, and energy density, do not show large discrepancies between the two scenarios in this temperature range, for all three values of $\mu_B/T$.
On the other hand, when realistic conditions of strangeness neutrality and fixed baryon-to-electric-charge densities are imposed on the global chemical composition of the system, the baryon density, for one, is largely affected.
In this case, it is substantially decreased, while for the electric charge density, the opposite effect is visible, as it is heavily enhanced.
Finally, the strangeness density is restricted to be exactly zero.
The behavior of the densities shows the substantial effect of the conditions that were imposed on the chemical potentials.

\begin{figure}[H]
    \centering
    \includegraphics[width=\linewidth]{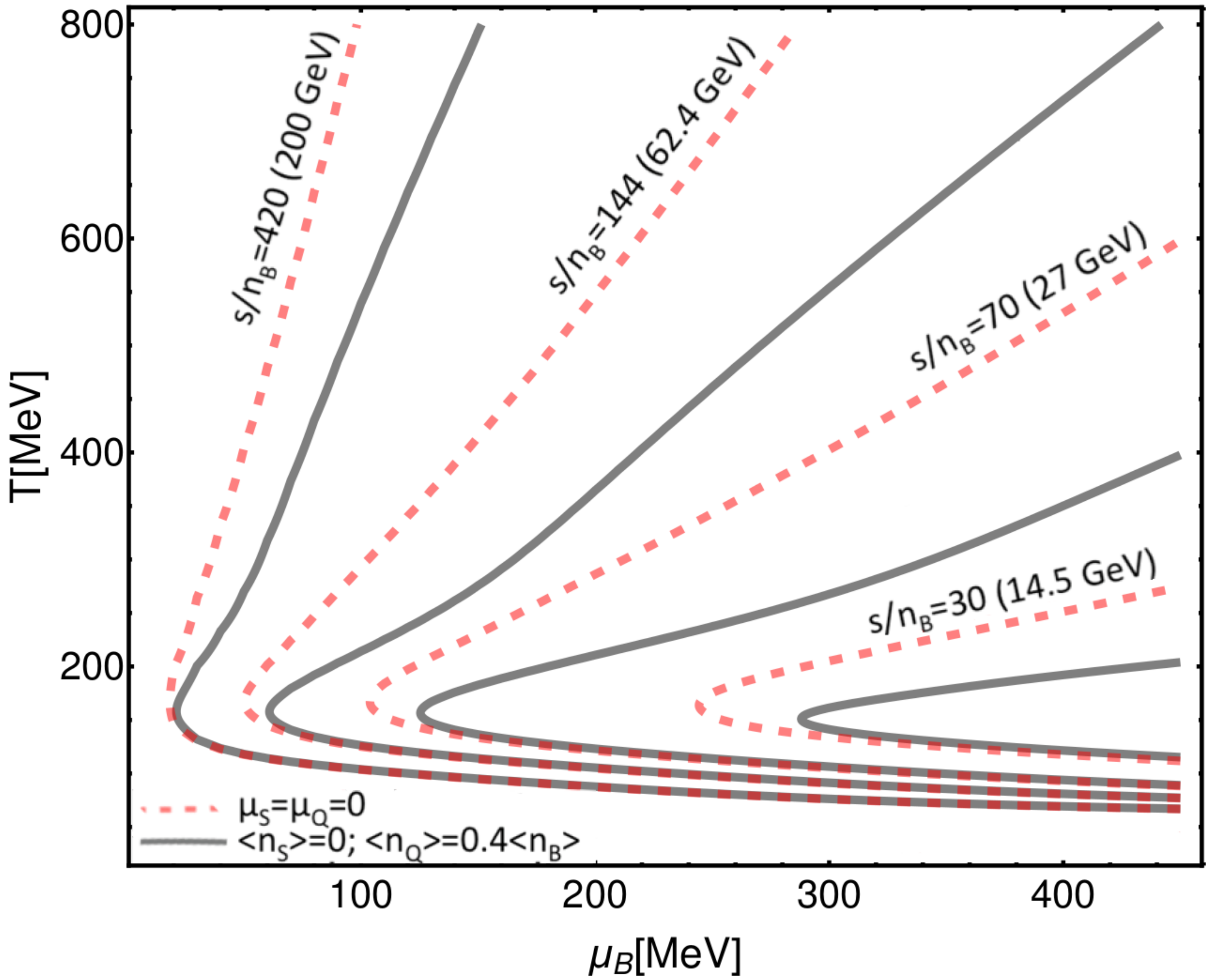}
    \caption{Isentropic trajectories in the $(T,~\mu_B)$ plane, for $s/n_B=420,~144,~70,~30$, corresponding to collision energies $\sqrt{s_{NN}}=200,~62.4,~27,~14.5$ GeV, respectively. The solid gray lines correspond to $\langle n_S\rangle=0$, $\langle n_Q\rangle=0.4\langle n_B\rangle$ while the dashed red lines to $\mu_S=\mu_Q=0$ \cite{Noronha-Hostler:2019ayj}.}
    \label{fig6}
\end{figure}

Furthermore, we make a comparison between these two cases of constraints on the chemical potentials for the isentropic trajectories.
The isentropic trajectories are shown in Fig. \ref{fig6} for selected values of $s/n_B$, which correspond to collision energies of $\sqrt{s_{\text{NN}}}=200,~62.4,~27,~14.5$ GeV \cite{Gunther:2016vcp}. 
For a fluid with a very small viscosity such as the QGP \cite{Schafer:2009dj,Bernhard:2016tnd}, these isentropic trajectories show the path of a heavy-ion-collision system through the phase diagram.
It is important to note that the system takes quite different paths through the phase diagram in the case of strangeness neutrality versus vanishing strangeness and electric charge chemical potentials.
As previously discussed, the entropy is not as strongly affected by the constraints on the conserved charge chemical potentials as the densities are.
This is the reason that the isentropic curves are pushed to larger values of $\mu_B$, corresponding to smaller values of the baryon number, in the case of strangeness neutrality for a given isentrope.
This is very impactful for the initial stages of HICs, which is another active research endeavor.

\section{BES EoS}

In order to study the effect of a critical point that could potentially be observed in the BES-II program at RHIC on QCD thermodynamics, we utilized the 3D Ising model to map such critical behavior onto the phase diagram of QCD.
The 3D Ising model was chosen for this approach because it exhibits the same scaling features in the vicinity of a critical point as QCD, in other words, they belong to the same universality class \cite{Pisarski:1983ms,Rajagopal:1992qz}.
The mapping of the critical behavior onto the QCD phase diagram is, however, not universal, which is to say that there are no strict mapping parameters that exist \textit{a priori} between the Ising phase diagram and the one for QCD.
We, thus, proceeded by implementing the non-universal mapping of the 3D Ising model in such a way that the Taylor expansion coefficients of our final pressure match the ones calculated from lattice QCD, order by order.
This ensures that our EoS matches the one from first-principles at $\mu_B=0$.
The prescription for implementing critical behavior into the QCD EoS based on universality arguments is summarized in the following step-wise procedure.
\begin{enumerate}
    \item \label{step:IsingEoS} Define a parametrization of the 3D Ising model near the critical point via the parameters $R$ and $\theta$, consistent with what has been previously shown in the literature  \cite{Guida:1996ep, Nonaka:2004pg,Stephanov:1998dy, Stephanov:2011pb, Parotto:2018pwx}:
    \begin{equation} \label{IsingEoS}
        \begin{split}
            M &= M_0 R^{\beta} \theta \\
            h &= h_0 R^{\beta \delta} \tilde{h}(\theta) \\
            t &= R(1- \theta^2).
        \end{split}
    \end{equation}
	The magnetization $M$, the  magnetic field $h$, and the reduced temperature $t$ along with the free energy make up the 3D Ising model equation of state. The critical scaling of the EoS is contained entirely in the behavior of $R$, in that it is the parameter for which the critical exponents appear. Those critical exponents in the 3D Ising model have the following values: $\alpha=0.11$, $\beta=0.326$, and $\delta=4.8$ \cite{Guida:1996ep}. Additionally, $R$ and $\theta$ can be understood as mapping the distance and angle displaced from the critical point, respectively. The normalization constants for the magnetization and magnetic field are $M_0=0.605$ and $h_0=0.364$, respectively. The magnetic field $h$ is proportional to a polynomial in odd powers of $\theta$: $\tilde{h}(\theta)=\theta (1+a\theta^2+b\theta^4)$, where $a=-0.76201, b=0.00804$. 
	
	The pressure in the Ising model is defined by the parametrized Gibbs' free energy up to a minus sign:
	\begin{equation}
	\label{GFreeEner}
	    \begin{split}
	     P_{\text{Ising}} &= - G(R,\theta) \\
	     &= h_0 M_0 R^{2 - \alpha}(\theta \tilde{h}(\theta) - g(\theta)),
	    \end{split}
	\end{equation}
	where
	\begin{align*}
	\centering
	     g(\theta) &= c_0 +  c_1(1-\theta^2) + c_2(1-\theta^2)^2 + c_3(1-\theta^2)^3, \\
        c_0 &= \frac{\beta}{2-\alpha}(1+a+b), \\
	 c_1 &= -\frac{1}{2} \frac{1}{\alpha -1}((1-2\beta)(1+a+b)-2\beta(a+2b)), \\
	 c_2 &= - \frac{1}{2\alpha}(2\beta b - (1-2\beta)(a+2b)), \\
	 c_3 &= - \frac{1}{2(\alpha+1)}b(1-2\beta).
    \end{align*}
	 
	This determines the singular part of the pressure, which carries the critical features, in this EoS mapped to QCD. Furthermore, because QCD is symmetric about $\mu_B=0$, we required that  $P_{\rm{Ising}}$ is also matter-antimatter symmetric. Thus, we performed the calculations in a range of $\mu_B$ spanning positive and negative values in order to symmetrize the pressure and all thermodynamic quantities. Additionally, the equations defined here were derived from a renormalization group approach and are subject to the following constraints on the parameters: R $\geq$ 0, $\lvert \theta \rvert$ $\leq$ $\theta_0$ $\sim$ 1.154, where $|\theta_0|$ correspond to the zeros of $\tilde{h}(\theta)$ \cite{Brezin:1976pt,Guida:1996ep}.
    \item \label{step:Isingmapping} Choose the location of the critical point and map the critical behavior onto the QCD phase diagram via a linear map from \{$T$, $\mu_B$\} to \{$t,h$\}:

    \begin{equation} \label{mapT}
        \frac{T-T_C}{T_C}=\omega(\rho t \sin{\alpha_1} + h \sin{\alpha_2})
    \end{equation}
    \begin{equation} \label{mapmuB}
        \frac{\mu_B-\mu_{B,C}}{T_C} = \omega(-\rho t \cos{\alpha_1} - h \cos{\alpha_2})
    \end{equation}
    
    where $T_C , \, \mu_{B,C}$ are the coordinates of the critical point, and $\alpha_1$,$\alpha_2$ are the angles between the axes of the QCD phase diagram and the Ising model ones. Finally, $\omega$ and $\rho$ determine the strength of the contribution of the Ising phase diagram in this mapping: $\omega$ determines the overall scale of both $t$ and $h$, while $\rho$ determines the relative scale between them.


    \item As previously established in Ref.~\cite{Parotto:2018pwx}, we reduced the number of free parameters from six to four, by assuming that the critical point sits along the chiral phase transition line, and by imposing that the $t$ axis of the Ising model is tangent to the transition line of QCD at the critical point: 
    \begin{equation} \label{chiraltrans}
        T=T_0 + \kappa \, T_0 \, \left(\frac{\mu_B}{T_0}\right)^2 + \mathcal{O}(\mu_B^4).
    \end{equation}
    In this study, we maintained consistency with the original EoS development of Ref.~\cite{Parotto:2018pwx} by utilizing the same parameters. As in Ref.~\cite{Parotto:2018pwx}, we assumed the transition line to be a parabola, and utilized the curvature parameter $\kappa=-0.0149$ from Ref.~\cite{Bellwied:2015rza}.
    Given this mapping, the critical point can be chosen to be anywhere along the chiral phase transition line. With this in mind, this equation of state is presented with a choice of critical point location relevant for the BES-II program. For this work, the critical point was mapped onto the QCD phase diagram at ($T_C \simeq 143.2$ MeV, $\mu_{B,C}=350$ MeV), the angular parameters are orthogonal $\alpha_1$=3.85\textdegree, $\alpha_2$=93.85\textdegree, and the critical-region-size parameters are $\omega$=1 and $\rho$=2.  However, it is imperative to note that this choice of parameters has only an illustrative purpose and that we do not make any statement about the position of the critical point or the size of the critical region. As this framework does not serve to yield a prediction for the critical point, but rather to provide an estimate of the effect of critical features on heavy-ion-collision systems, those using the open-source program are left to determine their preferred choice of parameters and test the effect on observables. In particular, we note that by varying the parameters $\omega$ and $\rho$ it is possible to  increase or decrease the effects of the critical point \cite{Parotto:2018pwx,Mroczek:2020rpm}.  Hopefully, experimental data from the BES-II will allow us to constrain the parameters and narrow down the location of the critical point. 
    
    \item Calculate the Ising model susceptibilities and match the Taylor expansion coefficients order by order to lattice QCD results at $\mu_B$=0. This construction ensures that our EoS reproduces the one from first-principles where available. The Taylor expansion of the pressure in $\mu_B$/T as calculated on the lattice can be written as:
    \begin{equation} \label{pressTaylor}
    \begin{split}
        \frac{P(T,\mu_B)}{T^4} = \sum_n c_{2n}(T) \left(\frac{\mu_B}{T} \right)^{2n}.
    \end{split}
    \end{equation}
    Thus, the background pressure, or the non-Ising pressure is, by construction, the difference between the lattice (LAT) and Ising contributions:
    \begin{equation} \label{coeffmatch}
        T^4 c_{2n}^{\rm{LAT}}(T) = T^4 c_{2n}^{\rm{Non-Ising}}(T) + T_C^4 c_{2n}^{\rm{Ising}}(T).
    \end{equation}
    \item The full Taylor-expanded pressure, including its critical and non-critical components, is calculated as:
    \begin{equation} \label{fullpress}
        P(T,\mu_B)=T^4 \sum_n c_{2n}^{\rm{Non-Ising}}(T) \left(\frac{\mu_B}{T} \right)^{2n}
        + P_{\rm{crit}}^{\rm{QCD}}(T,\mu_B),
    \end{equation}
    where $P_{\rm{crit}}^{\rm{QCD}} = T^4 P^{\text{Ising}}_{\text{symm}}$ is the critical contribution to the pressure that has been mapped onto the QCD phase diagram as described in steps \ref{step:IsingEoS} and \ref{step:Isingmapping}.
    
    \item Merge the full reconstructed pressure from the previous step with the HRG pressure at low temperature in order to smooth any non-physical artifacts of the Taylor expansion. For this smooth merging, we utilized a hyperbolic tangent:
    \begin{equation} \label{eq:Pmerging}
    \begin{split}
        \frac{P_{\text{Final}}(T,\mu_B)}{T^4} = \frac{P(T,\mu_B)}{T^4} \frac{1}{2}
        \Big[1 + \tanh{\Big(\frac{T-T'(\mu_B)}{\Delta T}}\Big)\Big] \\
        + \frac{P_{HRG}(T,\mu_B)}{T^4} \frac{1}{2}
        \Big[1 - \tanh{\Big(\frac{T-T'(\mu_B)}{\Delta T}}\Big)\Big],
    \end{split}
    \end{equation}
    where $T'(\mu_B)$ acts as the switching temperature and $\Delta T$ is the overlap region where both terms contribute. We performed the same merging as in the original development of the EoS, and therefore, utilized the functional form of the switching temperature to be parallel to the QCD transition line with an overlap region of $\Delta T$=17 MeV.
    \item Calculate thermodynamic quantities as derivatives of the pressure. Such quantities were obtained from the pressure according to the following relationships:
\begin{equation} \label{eq:thermodef-ch3}
\centering
    \begin{split}
         \,\,\,\, \frac{n_B}{T^3}&=\frac{1}{T^3} 
        \left( \frac{\partial P}{\partial \mu_B} \right)_T, \,\,\,\,\,\,\,\,\,
        \frac{\chi_2}{T^2}=\frac{1}{T^2} 
        \left( \frac{\partial^2 P}{\partial \mu_B^2} \right)_T, \\
        \frac{S}{T^3}&=\frac{1}{T^3} 
        \left( \frac{\partial P}{\partial T} \right)_{\mu_B}, \,\,\,\,\,\,\,\,\,
        \frac{\epsilon}{T^4}=\frac{S}{T^3} - \frac{P}{T^4} + \frac{\mu_B}{T} \frac{n_B}{T^3}, \\
        c_S^2 &= \left( \frac{\partial P}{\partial \epsilon} \right)_{S/n_B}.
    \end{split}
\end{equation}
\end{enumerate}

\begin{figure*}
    \centering
    \includegraphics[width=0.48\textwidth]{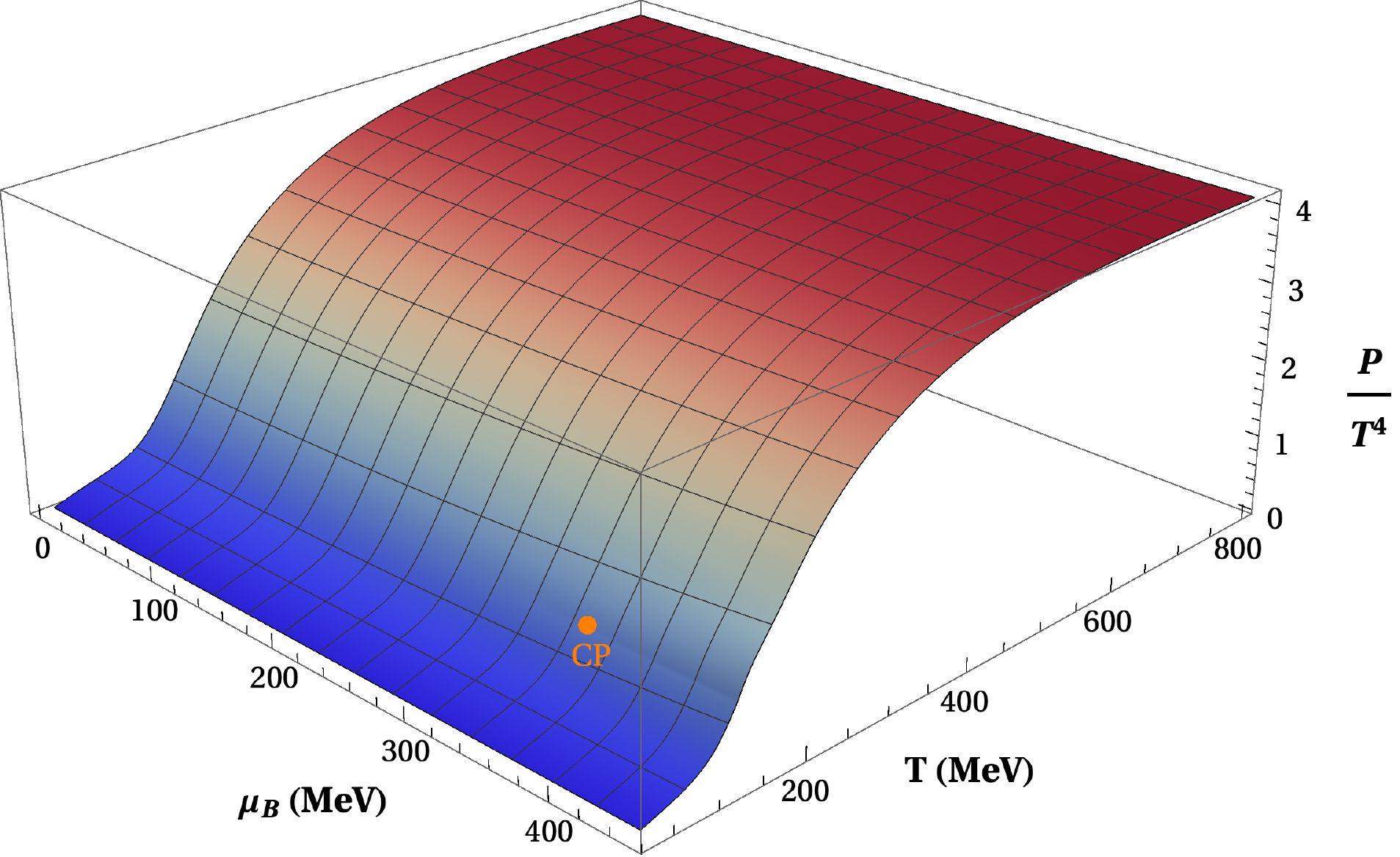}
    \includegraphics[width=0.5\textwidth]{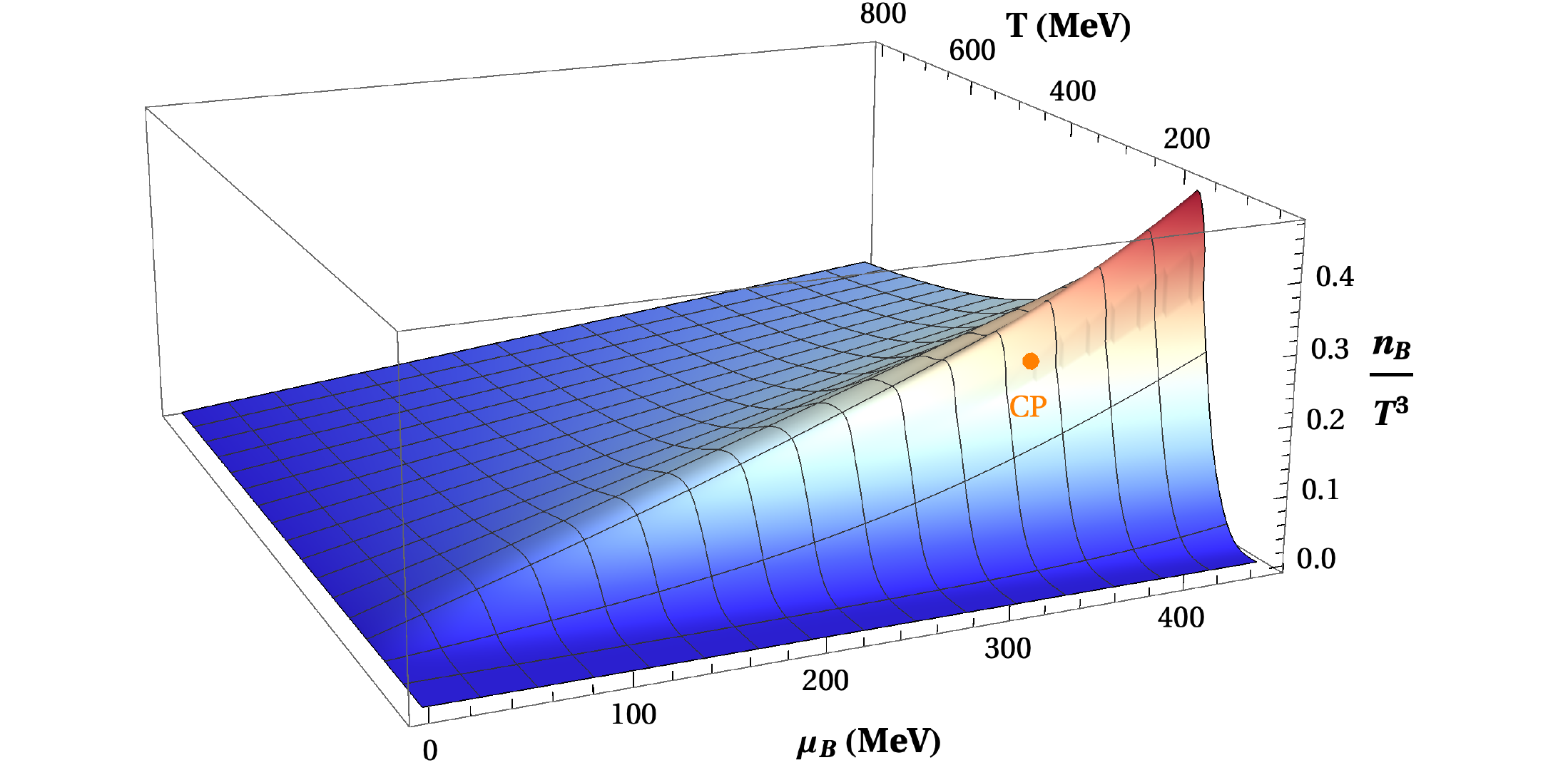}
    \caption{Left: The full QCD pressure for the choice of parameters consistent with Ref. \cite{Parotto:2018pwx}, as listed in the text.  Right: The baryon density for the same choice of parameters.}
    \label{fig:press}
\end{figure*}

\begin{figure*}
    \centering
    \includegraphics[width=0.49\textwidth]{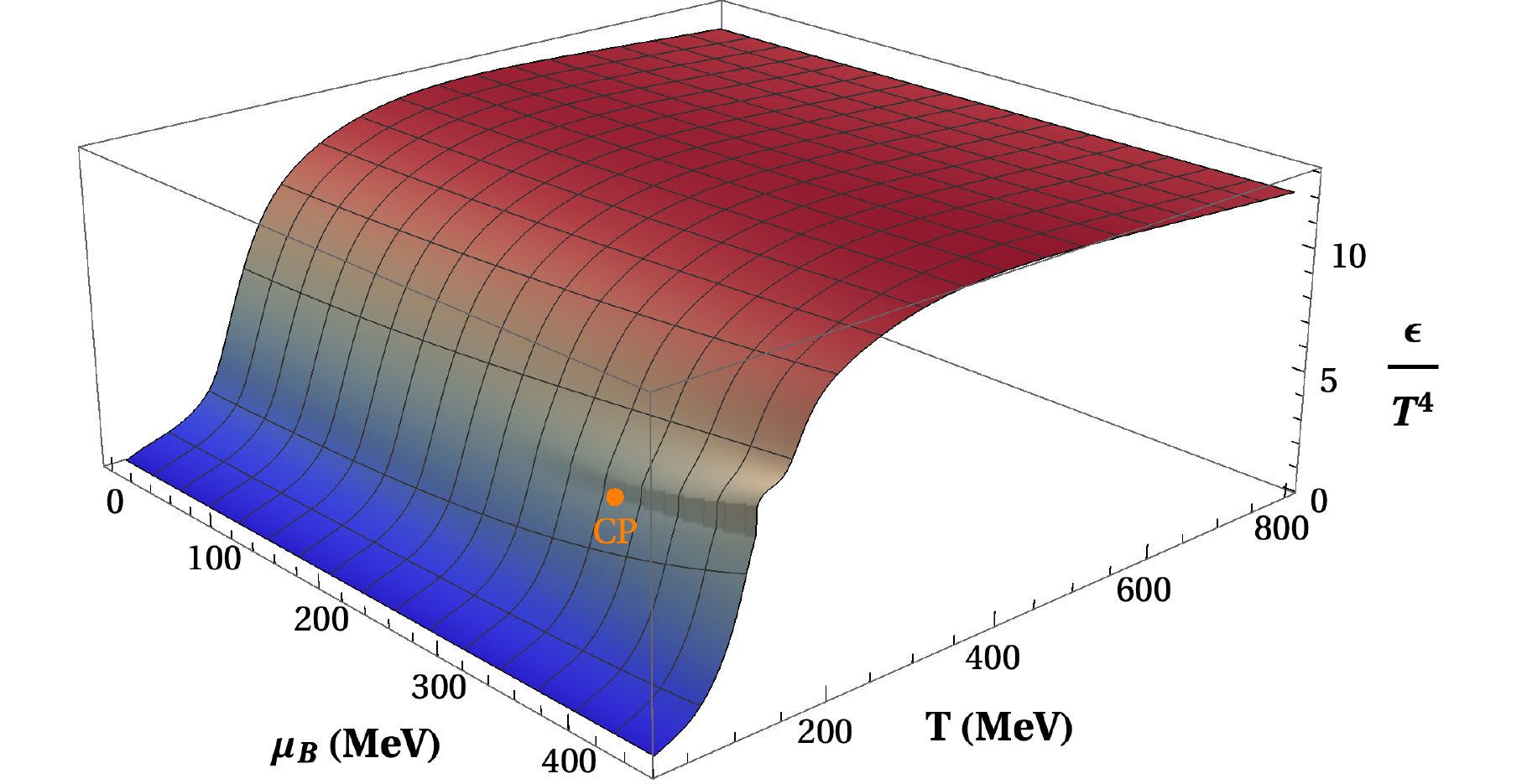}
    \includegraphics[width=0.49\textwidth]{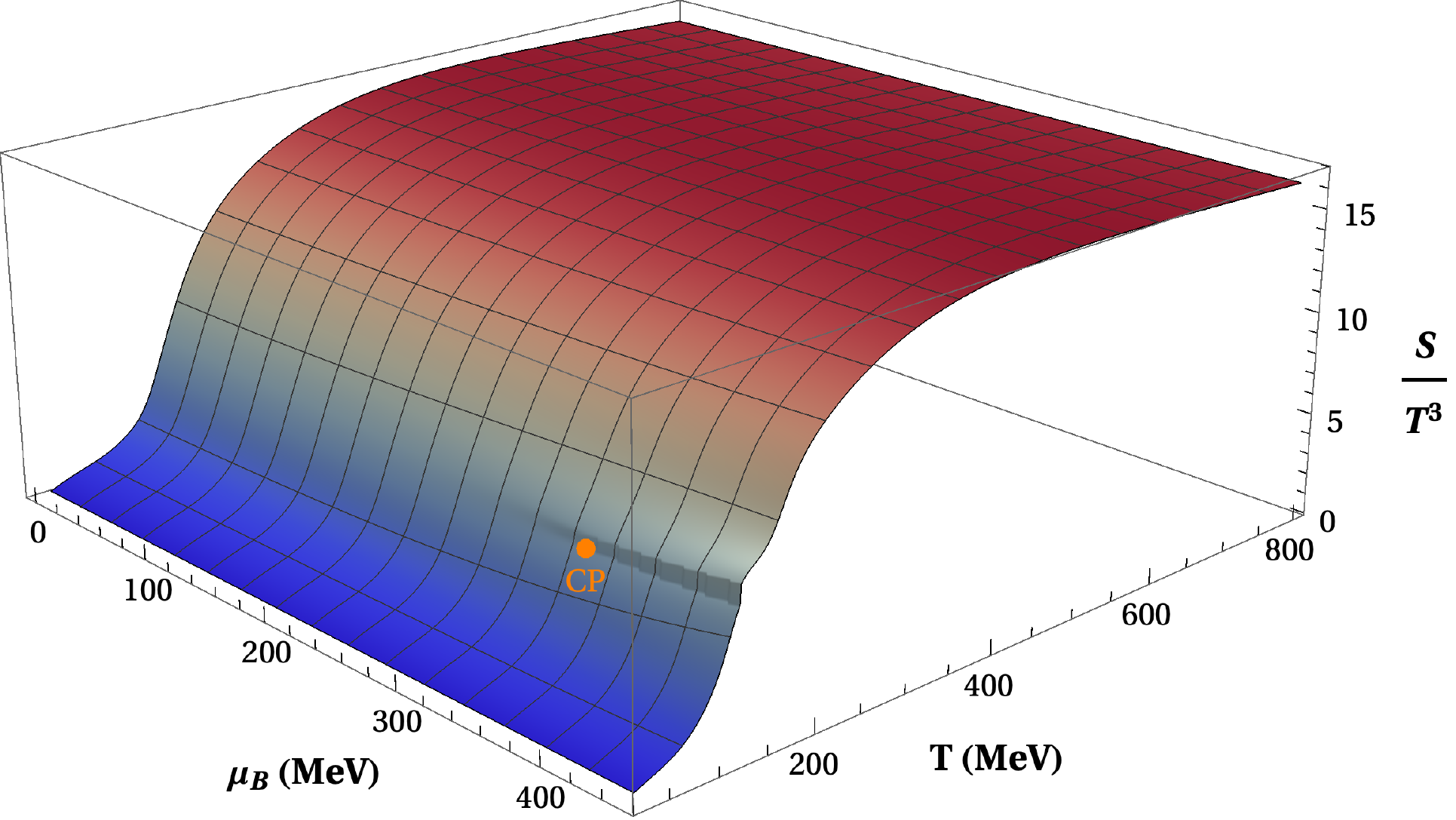}
    \caption{Left: The energy density for the choice of parameters consistent with Ref. \cite{Parotto:2018pwx}, as listed in the text. Right: The entropy density for the same choice of parameters.}
    \label{fig:enerdens}
\end{figure*}

\begin{figure*}
    \centering
    \includegraphics[width=0.6\textwidth]{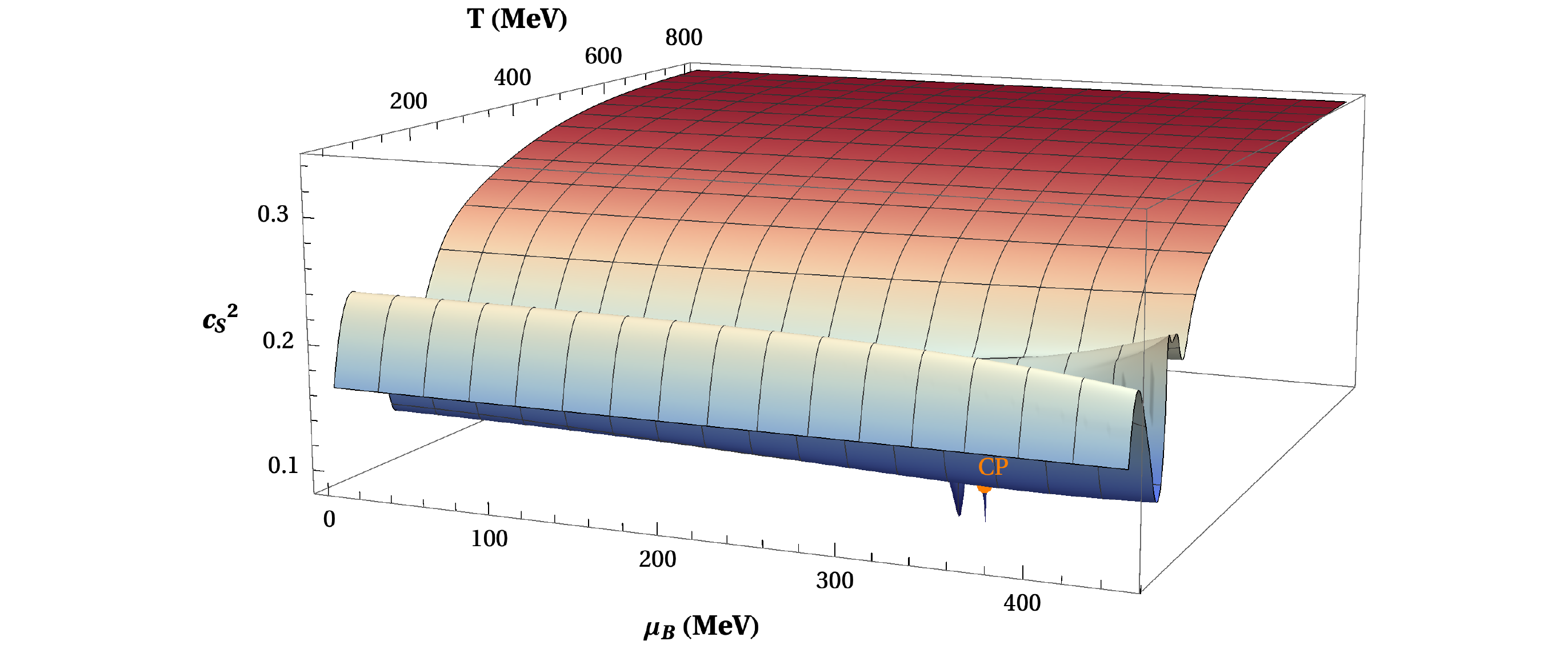}
    \caption{The speed of sound for the choice of parameters consistent with Ref. \cite{Parotto:2018pwx}, as listed in the text.}
    \label{fig:spsound}
\end{figure*}

The Equation of State created as a result of the procedure detailed here is shown in Figs. \ref{fig:press} - \ref{fig:spsound}.
The Taylor-expanded pressure is shown in the left panel of Fig. \ref{fig:press}, while the derivatives of the pressure (see Eq. \eqref{eq:thermodef-ch3} for definitions) are shown in the subsequent plots. In particular, the right panel of Fig. \ref{fig:press} shows the baryonic density, the two panels of Fig. \ref{fig:enerdens} show the energy density and entropy density, while Fig. \ref{fig:spsound} shows the speed of sound. 
These quantities are calculated via Eq. \eqref{eq:thermodef-ch3}.

The location of the critical point is indicated on each of these graphs to guide the reader to the critical region.
As expected, the pressure is a smooth function of \{T,  $\mu_B$\} in the crossover region with a slight kink for chemical potentials larger than $\mu_{B,C}$.
The derivatives of the pressure help to reveal the features of criticality, with an enhancement with increasing order of derivatives.
This intensification of critical features is due to the dependence on increasingly large powers of the correlation length, $\xi$.

\begin{figure*}
    \centering
    \includegraphics[width=0.5\textwidth]{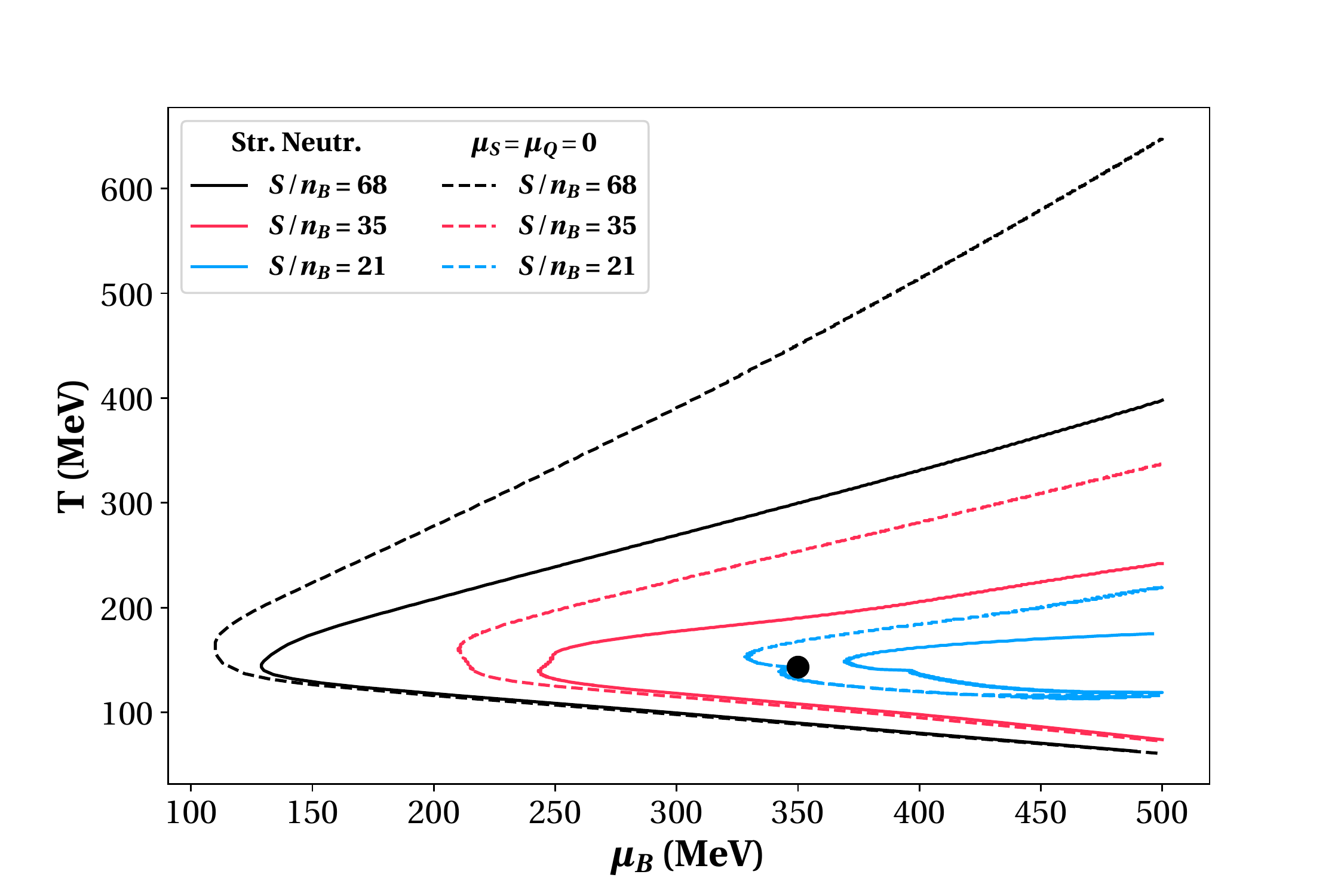}
    \caption{Comparison of the isentropes in the case of strangeness neutrality (dashed lines) and in the original formulation with $\mu_S=\mu_Q=0$ from Ref. \cite{Parotto:2018pwx} (solid lines). In both plots, the point marks the location of the critical point as listed in the text.}
    \label{fig:isen}
\end{figure*}

In Fig. \ref{fig:isen}, we present the isentropic trajectories in the QCD phase diagram for this framework, similarly to what is shown in Fig. \ref{fig6} for the BQS EoS.
If the isentrope passes through the critical region, it exhibits a disturbance where there is a jump in both the entropy and the baryon density. 
Fig.~\ref{fig:isen} shows the comparison between the new isentropes with conditions of strangeness neutrality and the previous ones from Ref. \cite{Parotto:2018pwx}.
While both exhibit features of the critical point, their trajectories through the phase diagram are quite different in the two cases, consistent with other EoS's that include strangeness neutrality conditions \cite{Noronha-Hostler:2019ayj,Monnai:2019hkn}.
Strangeness neutrality pushes the trajectories to larger $\mu_B$ for the same value of $T$, which was also seen in Fig. \ref{fig6}.
Since the isentropes can be understood to represent the path of the heavy-ion-collision system through the phase diagram in the absence of dissipation, we note that these plots, in particular, show the importance of incorporating strangeness neutrality into the EoS. 

\section{Conclusions}
The frameworks for equations of state that incorporate the relevant constraints on the conserved charge conditions for heavy-ion collision systems can readily be used in hydrodynamic simulations of heavy-ion collisions.
The programs for the EoSs can be found at the following repositories \cite{code:2019,BESTrepos}.
These conditions of strangeness neutrality and fixed baryon-number-to-electric-charge ratio were achieved in a purely first-principles-based approach and also for the phenomenologically motivated, criticality containing EoS from the BEST collaboration, the BES EoS. 
Not only do these EoSs probe a slice of the QCD phase diagram covered by the experiments, but they are also consistent with fundamental Lattice QCD results in the regions where they are available.
Furthermore, we have shown that the ideal trajectories through the phase diagram, as given by fixed values of entropy per baryon number, $s/n_B$, are vastly different in the case of strangeness neutrality as compared to vanishing strangeness and electric charge chemical potentials, which affects the initial conditions in HICs.

%
\nocite{*}
\bibliographystyle{rmf-style}
\bibliography{ref}

%
%
%

%
%

\section{Acknowledgements}

D.M. is supported by the National Science Foundation Graduate Research Fellowship Program under Grant No. DGE – 1746047 and the University of Illinois at Urbana-Champaign Sloan Graduate Fellowship. D.M. acknowledges support from the ICASU Graduate Fellowship. J.N.H. acknowledges financial support by the US-DOE Nuclear Science Grant No. DESC0020633. J.M.K. is supported by an Ascending Postdoctoral Scholar Fellowship from the National Science Foundation under Award No. 2138063. P.P. also acknowledges support by the DFG grant SFB/TR55.
C.R. acknowledges financial support
by the National Science Foundation under grant no. PHY-1654219.

\end{document}